\documentclass[pra,twocolumn,superscriptaddress]{revtex4-1}
\usepackage{float,graphicx,multirow, amsmath,color,dsfont}
\usepackage{amsmath,bm}
\usepackage{amssymb,mathrsfs,dsfont}
\usepackage{amssymb,mathbbol}
\usepackage{amsfonts}
\usepackage{mathrsfs}
\usepackage{float}
\usepackage{xcolor,hyperref}
\definecolor{darkblue}{rgb}{0,0.0.1,0.3}
\definecolor{darkred}{rgb}{0.6,0.1,0}
\hypersetup{colorlinks,breaklinks,
linkcolor=darkred,urlcolor=darkblue,
anchorcolor=darkred,citecolor=
darkred,pdfauthor=RiChGeAr, pdftitle=RiChGeAr.Dissipation}
\usepackage{tikz}
\usepackage{mathbbol}
\usepackage{float}
\newcommand{\ie}{\textit{i}.\textit{e}.}

\begin{document}
\title{Evolution of two-mode quantum states under a dissipative
environment: which quantum resource survives better, 
squeezing or entanglement?}
\author{Rishabh}
\email{rishabh1@ucalgary.ca}
\altaffiliation[Present Address~:~]{Department of Physics and
Astronomy, University of Calgary, Calgary
T2N1N4, Alberta, Canada.}
\affiliation{Department of Physical Sciences, Indian
Institute of Science Education and Research (IISER), Mohali,
Sector 81 SAS Nagar, Manauli PO 140306, Punjab, India}
\author{Chandan Kumar}
\email{chandan.quantum@gmail.com}
\affiliation{Department of Physical Sciences, Indian
Institute of Science Education and Research (IISER), Mohali,
Sector 81 SAS Nagar, Manauli PO 140306, Punjab, India}
\author{Geetu Narang}
\email{geet29@gmail.com}
\affiliation{Department of Applied Sciences, U.I.E.T, Panjab
University, Chandigarh 160040, India}
\author{Arvind}
\email{arvind@iisermohali.ac.in}
\affiliation{Department of Physical Sciences, Indian
Institute of Science Education and Research (IISER), Mohali,
Sector 81 SAS Nagar, Manauli PO 140306, Punjab, India}
\affiliation{Vice Chancellor, Punjabi University Patiala,
Punjab 147002, India}
\begin{abstract}
We explore the relative robustness of squeezing and
entanglement (which are quantum resources interconvertible
via passive optics) for  two-mode Gaussian  states under
different dissipative environments.  When the individual
modes interact with identical local baths, entanglement and
squeezing decay at the same rate.  However, when only one of
the modes interacts with a local bath, the comparative
robustness of entanglement and squeezing depends on the
initial squeezing of the state.  Similarly, when the system
interacts with a global bath, the robustness of entanglement
and squeezing depends on the initial squeezing.  Thus
depending on the nature of dissipative environments and the
initial squeezing of the state, one can select the more
robust form of resource out of squeezing and entanglement to
store quantumness. This can be used to  effectively enhance
the performance of various quantum information processing
protocols based on continuous variable Gaussian states.
\end{abstract}
\maketitle
\section{Introduction}
\label{sec:intro}
Continuous variable (CV) quantum information processing
(QIP) based on quantum optical sources has been  gaining
attention over time~\cite{weedbrook-rmp-2012,adesso-2014}.
Developments in continuous variable quantum key
distribution(CV-QKD) protocols~\cite{pirnadola-2019} and
photonic computing~\cite{Ish-2021} are particularly
noteworthy.  Nonclassical Gaussian states play a key role in
this context, as they can be easily produced, manipulated,
and measured in the
laboratory~\cite{10db-prl-2008,15db-prl-2016}.
Nonclassicality or genuine quantumness in quantum states is
a resource that needs to be defined, identified, and
preserved for use in QIP
protocols~\cite{simon-prl-2000,duan-prl-2000,pablo-prl-2008,nonclassicality-prx-2018}.

In the quantum optical sense, if the Glauber-Sudarshan $P$
function~\cite{Glauber,Sudarshan} behaves like a regular
probability distribution, the corresponding state can be
simulated by  ensembles of solutions of Maxwell equations
and the state cannot exhibit any nonclassical
features~\cite{mandel_wolf_1995}.  On the other hand,
nonpositivity of the Glauber-Sudarshan $P$ function
indicates nonclassicality~\cite{Adam-pra-2010,Nori-pra-2011}.
Squeezing, when quadrature noise drops below the shot noise limit,
is one specific form of quantumness based on P
representation~\cite{gerry_knight_2004}. 
Gaussian squeezed  states are thus an important class
of nonclassical states which are extremely useful for QIP.
A different notion of quantumness arises from   information
theory viewpoint, where correlations in composite quantum
systems can go beyond classically allowed values leading to
nonclassical situations.  Quantum entanglement is one such
resource that can lead to situations that violate local
realism~\cite{horodecki-rmp-2009}.  While single mode
squeezing and intermode entanglement are very different
notions of nonclassicality, they can be inter-converted into
each other via passive optical elements such as beam
splitters, phase shifters, wave-plates, and
mirrors~\cite{scheel-pra-2001,xiang-pra-2002,wolf-prl-2003,Bowen_2003}.

Environmental interactions can cause disturbances which
invariably lead to the diminishing of quantum resources which
may be present in the form of squeezing and entanglement.
The environmental interactions are detrimental to the
performance of various QIP tasks. Therefore, it is of
significant importance to analyze the evolution of quantum
systems under different dissipative environments and find
ways to protect resources against environmental effects. A
lot of work has already been done in this
regard~\cite{zhao-pra-2002,prauzner-2004,serafini-2005,pablo-prl-2008,xiang-pra-2008,xiang-pra-2009,sandeep-pra-2010,xiang-epjd-2011}.
Entanglement dynamics  has been studied for local as well as
global dissipative environments under both Markovian and
non-Markovian assumptions.  Many interesting phenomena have
been observed, for instance, sudden death of entanglement in
local as well as global dissipative environments~\cite{pablo-prl-2008, sandeep-pra-2010}.
 Further, researchers have also shown that entanglement can be
produced in two-mode separable squeezed states and can even
be enhanced in two-mode squeezed vacuum (TMSV) states
evolving under the presence of a global thermal
bath~\cite{benatti-2006,xiang-pra-2008}.  The effect of
decoherence on nonlocality, discord, and steering has also
been
studied~\cite{liang-pra-2011,pablo-pra-2012,daffer-pra-2003,wang-qip-2016}.
Moreover, much research has been conducted to study
decoherence in non-Gaussian states, particularly photon
subtracted states~\cite{agarwal-pra-2007,hu-pra-2010}.

In this article, we strive to find which of the two
resources, squeezing or entanglement, is more robust to
environmental
noise~\cite{geetu-aip-2006,Serafini-pla-2011,Solomon-pra-2019}.
We consider two different cases that provide an insight into
the relative sensitivity of squeezing and entanglement
resources to dissipative environments  under the  Markovian
assumption.  The first case considers squeezing of the
individual modes followed by an evolution under a
dissipative environment, and finally, the two modes are
entangled using passive optics (beam splitter).  The second
case considers squeezing of the individual modes, which then
are entangled  using passive optics, and finally, we let
them evolve in a dissipative environment.  We have
considered our system to be interacting with local and
global thermal baths, which are Gaussian
channels, \ie,  Gaussian states remain Gaussian under such
interactions. We provide
the time-dependent covariance matrix, which is used for
entanglement analysis.  
Although a full characterization of entanglement for
CV systems is not possible, for Gaussian states,
necessary and sufficient criteria  for the detection of
entanglement exist~\cite{simon-prl-2000,duan-prl-2000}.
Further, we can quantify entanglement using logarithmic
negativity in two-mode Gaussian
states which we use in our 
work~\cite{lewenstein-pra-1998,vidal-pra-2002,plenio-prl-2005}.

It is natural to expect that entanglement in the TMSV state,
where intermodal correlations are present, will be more
fragile compared to squeezing in a two-mode separable
squeezed state.  However, our analysis reveals that the
relative robustness of entanglement and squeezing   depends
on the dissipative environment that the system is
interacting with.
When the two modes interact with identical baths, the
squeezing and entanglement decay in exactly the same way.
On the other hand, when only one of the modes interacts with
a local bath, the results depend on the initial squeezing
of the state.  There exists a threshold of the initial
squeezing of the state,  below which entanglement is more
robust than squeezing. Otherwise, squeezing is more robust
than entanglement. Similarly, such a threshold also exists
when the system interacts with a global bath, below which
squeezing is more robust than entanglement.  Otherwise,
entanglement is more robust than squeezing.  We provide
analytical expressions for these thresholds of the squeezing
parameter, which will enable experimentalists to identify
more robust resources against a dissipative environment. 

The paper is organized as follows. In Sec.~\ref{review}, we
review the formalism for the CV systems and describe the
notions squeezing and entanglement for two-mode Gaussian
states. In Sec.~\ref{resultsintro}, we set out to study the
relative robustness of squeezing and entanglement against a
disturbance caused by a noisy dissipative environment. We
consider the system evolution under local thermal baths in
Sec.~\ref{resultslocal}, while Sec.~\ref{resultsglobal}
deals with the evolution of the system under a global bath.
Finally, in Sec.~\ref{conc}, we provide some concluding
remarks and future directions.
%%%%%%%%%%%%%%%%%%%%%%%%%%%%%%%%%%%%%%%%%%%%%%%%
\section{Two mode systems, entanglement and squeezing}
\label{review}
%%%%%%%%%%%%%%%%%%%%%%%%%%%%%%%%%%%%%%%%%%%%%%%%
We describe the formalism of two-mode CV systems and discuss the
two nonclassical notions namely squeezing and entanglement
which we intend to study under different noisy dissipative
environments.
\subsection{Two mode CV systems and symplectic
transformations}
%%%%%%%%%%%%%%%%%%%%%%%%%%%%%%%%%%%%%%%%%%%%%%%
We consider a two-mode CV quantum system described  by the  
quadrature operators $\hat{q}_1$, $\hat{p}_1$, $\hat{q}_2$,
and $\hat{p}_2$~\cite{arvind1995,Braunstein,weedbrook-rmp-2012,adesso-2014,arvind1995}.
To handle the analysis of the two-mode system compactly, we
introduce the column vector
\begin{equation}\label{eq:columreal}
\bm{\hat{ \xi}} =(\hat{ \xi}_j)= (\hat{q}_{1},\,
\hat{p}_{1},\,\hat{q}_{2}, 
\, \hat{p}_{2})^{T}.
\end{equation}
The canonical commutation relations can be written as
\begin{equation}
\begin{array}{c}
[\hat{\xi}_j, \hat{\xi}_k] = 
i (\omega^{\oplus 2})_{jk}\\ 
{\scriptstyle j,k=1,2,3,4}\\
{\scriptstyle \hbar=1}
\end{array}\quad
{\rm with}\quad
\omega = \begin{pmatrix}0 & 1\\ -1 & 0\end{pmatrix}.
\label{eq:ccr}
\end{equation}
The  annihilation and creation 
operators $\hat{a}_j\, \text{and}\, {\hat{a}_j}
^{\dagger}$ ($j=1,2$) 
can be expressed in terms of quadrature operators as follows:
\begin{equation}\label{realtocom}
\hat{a}_j=\frac{1}{\sqrt{2}}(\hat{q}_j+i\hat{p}_j),
\quad \hat{a}^{\dagger}_j= \frac{1}{\sqrt{2}}(\hat{q}_j-i\hat{p}_j).
\end{equation}
The  Hilbert space of the two-mode system has an
orthogonal basis in the Fock representation $ \vert n_1 ,
n_2\rangle$ with $\{n_1,\, n_2=0,\, 1, \dots ,\infty \} $,
which are simultaneous eigenvectors of the number operators
$\hat{a}_1^{\dagger} \hat{a}_1$ and  $\hat{a}_2^{\dagger}
\hat{a}_2$.

The linear homogeneous transformations  $S$ specified by
real $4 \times 4$ matrices acting on the quadrature
operators~(\ref{eq:columreal}) and preserving the  canonical
commutation relation~(\ref{eq:ccr}) form  the symplectic
group $Sp(4,\,\mathcal{R})$. The quadrature operators
transform as $\hat{\xi}_i \rightarrow \hat{\xi}_i^{\prime} =
S_{ij}\hat{\xi}_{j}$.  Further, the symplectic condition for
matrix $S$ is
\begin{equation} 
S \omega^{\oplus 2} S^T = \omega^{\oplus 2} \implies S \in Sp(4,\,\mathcal{R}).  
\end{equation}
The symplectic group can be decomposed as $S= P K(X,Y)$,
where $P$ is the noncompact part and $K(X,Y)$ is the
maximally compact subgroup of $Sp(4,\mathcal{R})$.  The
elements of the set $P$ act on the states through their
infinite-dimensional unitary representation (metaplectic
representation), change the total number of photons and
are called active operations. Active operations can
transform a classical state into a nonclassical state and
vice versa. On the other hand, the elements of $K(X,Y)$
which form the maximally compact subgroup while
acting on the states via their metaplectic representation
conserve the total number of photons and are called passive
operations. Passive operations cannot create or
destroy the nonclassicality of a state and can be implemented
using passive optical elements like beam splitters, phase
shifters, waveplates, and mirrors. We discuss two basic symplectic
operations which are important for our 
work~\cite{arvind1995,Braunstein,weedbrook-rmp-2012,adesso-2014}.
\par
\noindent
{\bf Single mode squeezing operation~:}
The  transformation matrix corresponding to the 
single-mode squeezing operator, which acts on the 
quadrature operators $\hat{q}_i$ and $\hat{p}_i$, is given by
\begin{equation}\label{squeezer}
S_i(r) = \begin{pmatrix}
e^{-r} & 0 \\
0 & e^{r}
\end{pmatrix}.
\end{equation}
The corresponding $Sp(4,\mathcal{R})$ element will be
$S_1(r)\oplus \mathbb{1}_2$ or $\mathbb{1}_2\oplus S_2(r)$
depending upon which
of the two modes the squeezing operator acts upon.
Here $\mathbb{1}_2$ represents a $2 \times 2$ identity matrix.
Single-mode squeezing operators are active operations which can
transform a classical state into a nonclassical state.

\par
\noindent
{\bf Beam splitter operation\,:}
The  transformation matrix corresponding to a beam splitter 
acting  on the two-mode quadrature operators $  \hat{\xi} =
(\hat{q}_{1}, \,\hat{p}_{1},\,
\hat{q}_{2},\,\hat{p}_{2})^{T}$
is given by
\begin{equation}\label{beamsplitter}
B_{12}(\theta) = \begin{pmatrix}
\cos \theta \,\mathbb{1}_2& \sin \theta \,\mathbb{1}_2 \\
-\sin \theta \,\mathbb{1}_2& \cos \theta \,\mathbb{1}_2
\end{pmatrix},
\end{equation}
where 
$\theta$ is related to the  transmittance of the beam
splitter according to  $\tau = \cos ^2 \theta$. 
Angle $\theta = \pi/4$ corresponds to a balanced (50:50)
beam splitter. Beam splitter transformation is a passive operation
belonging to the $K(X,Y)$ subgroup of $Sp(4,\mathcal{R})$ which
cannot transform the classical or nonclassical status of a
state.
%%%%%%%%%%%%%%%%%%%%%%%%%%%%%%%%%%%%%%%%%%%%%%%%%%%%%%%
\subsection{Squeezing and entanglement in Gaussian states} 
Gaussian states can be described by Gaussian-Wigner
functions in phase space. They can be completely specified
by the mean values and covariances of the quadrature
operators. However, without any loss of generality, we can
consider a zero-centered state as any general Gaussian state
can be made zero-centered by application of the displacement
operator without affecting the entanglement content of the
state. The Wigner function for zero-centered Gaussian states
can be written
as~\cite{weedbrook-rmp-2012}
\begin{equation}
\label{eq:wignercovariance}
W(\bm{\xi}) = \frac{\exp[-(1/2)\bm{\xi}^TV^{-1}
\bm{\xi}]}{(2 \pi)^2 \sqrt{\text{det}V}},
\end{equation}
where $V$ is the covariance matrix whose elements
are given by
\begin{equation}\label{eq:cov}
V = (V_{ij})=\frac{1}{2}\langle \{\Delta \hat{\xi}_i,\Delta
\hat{\xi}_j\} \rangle,
\end{equation}
where $\Delta \hat{\xi}_i = \hat{\xi}_i-\langle \hat{\xi}_i
\rangle$, and $\{\,, \, \}$ denotes the anti-commutator.
The uncertainty principle can be expressed in terms of
the covariance matrix as
\begin{equation}
V+\frac{i}{2}\Omega \geq 0.
\end{equation}
The covariance matrix for the zero-centered two-mode Gaussian
state is given by
\begin{equation}\label{eq:4}
V=\left(\begin{array}{cccc}
\langle q_1^2\rangle&\frac{1}{2}\langle \{q_1,p_1\}\rangle&
\langle q_1 q_2\rangle  &
\langle q_1 p_2\rangle \\ 
\frac{1}{2}\langle \{q_1,p_1\}\rangle &
\langle p_1^2\rangle&
\langle q_2 p_1 \rangle& \langle p_1 p_2 \rangle \\
\langle q_1 q_2\rangle &\langle q_2 p_1\rangle&
\langle q_2^2\rangle  &
\frac{1}{2}\langle \{q_2,p_2\}\rangle\\
\langle q_1 p_2\rangle&\langle p_1 p_2 \rangle
&
\frac{1}{2}\langle \{q_2,p_2\}\rangle & \langle p_2^2
\rangle
\end{array} \right).
\end{equation}
A quantum state is said to be squeezed in quadrature
$\hat{\xi}_i$ if the fluctuations in the corresponding
quadrature reduce below the coherent state value, \ie,
$(\Delta \hat{\xi}_i)^2 <1/2$. The single mode squeezing
operator $S(r)$~(\ref{squeezer}) acting on a state through
its metaplectic representation alters fluctuations in the
quadratures, and hence, can transform a non-squeezed state
into a squeezed state. Specifically, as an example, if
$S(r)$ acts on the first mode, the fluctuations in the
$\hat{\xi}_1$ quadrature transform as $(\Delta
\hat{\xi}_1)^2 \rightarrow e^{-2r}(\Delta \hat{\xi}_1)^2$.
Since one is allowed to redefine quadratures by  mixing them
via passive operations, a state is squeezed even if the
noise in one of the transformed quadratures falls below the
coherent state value.  We will consider squeezing caused by
single mode squeezing transformations $S(r)$ as described
above.

While the detection of
entanglement in general states of CV systems remains an open
problem, for Gaussian states, the Simon criterion provides
necessary and sufficient conditions for the detection of
entanglement~\cite{duan-prl-2000,simon-prl-2000}.  By
Simon's criterion, 
\begin{equation}\label{simon_criterion}
\begin{aligned}
\, &\text{det}\,A\,\text{det}\,B +
\Big{(}\frac{1}{4}-|\text{det}\,C|\Big{)}^{2}-\text{Tr}[A\omega
C\omega B\omega C^{T}\omega]\\
&-\frac{1}{4}(\text{det}\,A+\text{det}\,B) \geq 0,
\end{aligned}
\end{equation}
is necessary and sufficient for a state to be separable.
Here $A$, $B$, and $C$ are $2\times 2$ matrices and are
related to the covariance matrix of a two-mode system as
\begin{equation}\label{Cov.matrix-ABC}
V=\left(\begin{array}{cccc}
A &C\\
C^{T} & B\end{array} \right).
\end{equation}
Furthermore for Gaussian states, the logarithmic negativity can
be used as a measure of
entanglement between the two modes. The logarithmic 
negativity for  a two-mode Gaussian state is defined as
\begin{equation}\label{eq:15}
E_{N} = \text{max}\{0,-\text{log}_{2}(2n_{-})\},
\end{equation}
where $n_{-}$ is the smallest symplectic eigenvalue of 
the partially transposed covariance matrix, which can be
compactly written using the block matrices form of the covariance
matrix~(\ref{Cov.matrix-ABC}) as follows:
\begin{equation}\label{eq:16}
\begin{aligned}
n_{-}^2 = &\frac{1}{2}\bigg[\Sigma
-\sqrt{\Sigma^2-4\,\text{det}\,{V}}\bigg],
\end{aligned}
\end{equation}
where $\Sigma = \text{det}\,A + \text{det}\,B -
2\,\text{det}\,C$.
We employ this measure to quantify entanglement throughout
this paper.
\section{Effect of environment on quantum resources of
squeezing and entanglement}
\label{results}
In this section, we set out to explore the  effect of
environmental coupling on diminishing the quantum resources
of squeezing and entanglement. We couple the two-mode system
with thermal baths in several different ways to study the
relative performance of squeezing and entanglement in terms
of their ability to withstand the effects of environmental
effects. 
%%%%%%%%%%%%%%%%%%%%%%%%%%%%
\subsection{Environment coupled to the two-mode optical system}
\label{resultsintro}
We consider a two-mode system interacting with a thermal
bath. The bath is considered to be comprised of
a large set of harmonic oscillators.  The corresponding
total Hamiltonian can be written as
\begin{equation}\label{hamiltonian}
\hat{H} = \hat{H}_{S} + \hat{H}_{B} + \hat{H}_{SB},
\end{equation}
where $\hat{H}_S$ and $\hat{H}_B$ are the system and the
bath Hamiltonians, respectively, and $\hat{H}_{SB}$ is the
interaction Hamiltonian. 
We consider two distinct cases:
\par
\noindent
\textbf{Case 1:} In this case a two-mode separable
squeezed state interacting with a thermal bath is considered. 
We start with a two-mode system initialized to the vacuum state 
$|\textbf{0}\rangle$, which can be represented by the
following covariance matrix:
\begin{equation}\label{vacuum}
V_{|\textbf{0}\rangle}=\frac{1}{2}\left(\begin{array}{cccc}
1 & 0 & 0 & 0\\
0 & 1 & 0 & 0\\
0 & 0 & 1& 0\\
0 & 0 & 0 & 1\end{array} \right).
\end{equation}
To generate a two-mode separable squeezed state, 
the two-mode vacuum state is squeezed by equal and
opposite amounts by the squeezing operation $S_1(r)\oplus
S_2(-r)$. Thereafter, the covariance matrix of the two-mode
separable squeezed  state can be written as
\begin{equation}\label{initial_V_case1}
V_{1}(t=0)=\frac{1}{2}\left(\begin{array}{cccc}
e^{-2r} & 0 & 0 & 0\\
0 & e^{2r} & 0 & 0\\
0 & 0 & e^{2r} & 0\\
0 & 0 & 0 & e^{-2r}\end{array} \right).
\end{equation}
The system is then allowed to interact with a thermal bath.  
We set the time to $t=0$, when the interaction with the bath
is switched on. Thus, the covariance matrix of the 
state at time $t=0$ is given by 
Eq.~(\ref{initial_V_case1}).
The interaction with the bath thermal bath is then switched
on which can lead to decay of squeezing.
After a time $t=\tau$ ,
the interaction is switched off, and the modes are 
mixed using a 50:50 beam splitter  with a view to 
convert the remaining nonclassicality into entanglement.
\par
\noindent
\textbf{Case 2:} In this case, we consider a two-mode squeezed 
vacuum state (TMSV), an entangled state, interacting 
with a thermal bath. The two-mode entangled state is
generated by first squeezing the two-mode system in the vacuum
state by an equal and opposite 
amount by the squeezing operation $S_1(r)\oplus S_2(-r)$.
The modes are then mixed via a 
50:50 beam splitter in order to convert the squeezing into
entanglement, and subsequently the thermal bath is switched on. The
covariance matrix at time $t=0$ for this case is given by 
\begin{equation}\label{initial_V_case2}
V_{2}(t=0) = \frac{1}{2}\begin{pmatrix}
\cosh (2r)\, \mathbb{1}_{2}& \sinh (2r) \,\mathbb{Z} \\
\sinh (2r) \,\mathbb{Z}& \cosh (2r) \,\mathbb{1}_{2}
\end{pmatrix},
\end{equation}
where $\mathbb{Z}$ is $\text{diag}(1,-1)$.
It should be noted here that for the two-mode separable squeezed
state, the quantum resource is in the form of squeezing,
while for the TMSV
state, squeezing has been converted into
entanglement~\cite{zubairy-pra-2009,zubairy-pra-2015}. Thus,
for a two-mode mode separable squeezed state, squeezing decays
due to environmental interactions.  On the other hand, for
TMSV state, both squeezing and entanglement decay due to
environmental interactions. The two cases differ from each
other only in the fact that  quantumness has been
completely stored in the form of squeezing in the former case,
while in the latter case, quantumness has been
converted  into the form of entanglement. The settings
above have been constructed to address
our main question: which form of quantumness,
squeezing or entanglement, is more resilient to a dissipative
environment? To this end, we consider the evolution of the
two aforementioned cases in the following two 
dissipative environments:
\begin{itemize}
\item[1.] The two modes of the system interact with two local
thermal baths.
\item[2.] The two modes of the system interact with a
global thermal bath
\end{itemize}
These studies will enable us to answer whether we should 
store the quantum resource as squeezing or as entanglement
in a given situation.
%%%%%%%%%%%%%%%%%%%%%%%%%%%%%%%%%%%%%%%%%%%%%%%%%%%%%%%
\subsection{Evolution under local thermal baths}
\label{resultslocal}
In this subsection, we consider that our two-mode system 
interacts with two local thermal baths. The interaction
Hamiltonian is given by
\begin{equation}\label{int_hamiltonian_local}
\hat{H}_{SB} =
g_{1}\sum^{\infty}_{k=1}\Big{(}\hat{a}_{1}\hat{b}^{\dagger}_{k}
+ \hat{a}^{\dagger}_{1}\hat{b}_{k}\Big{)}
+g_{2}\sum^{\infty}_{l=1}\Big{(}\hat{a}_{2}\hat{c}^{\dagger}_{l}
+ \hat{a}^{\dagger}_{2}\hat{c}_{l}\Big{)},
\end{equation}
where $g_{1}$ and $g_{2}$ are the coupling constants, and
$\hat{b}_k$ and $\hat{b}^{\dagger}_k$ are annihilation and
creation operators of the $k^{th}$ mode of the reservoir 
interacting with the first mode of the system. Similarly,
$\hat{c}_l$ and $\hat{c}^{\dagger}_l$ are annihilation and
creation operators of the $l^{th}$ mode of the reservoir 
interacting with the second mode of the system.

Under Markovian assumption, we can write the master 
equation for the evolution of system density operator $\rho$ as
\begin{equation}\label{master_eqn_local}
\begin{aligned}
\frac{\partial}{\partial t} \rho = & \bigg\{ \sum_{i=1,2}
\frac{\gamma_{i}}{2}(N_{i}+1)(2 \hat{a}_i \rho
\hat{a}_i^{\dagger}-\hat{a}_i^{\dagger}\hat{a}_i\rho -\rho
\hat{a}_i^\dagger\hat{a}_i)\\
& \quad + \frac{\gamma_{i}}{2}N_{i}(2 \hat{a}_i^\dagger \rho
\hat{a}_i-\hat{a}_i\hat{a}_i^{\dagger}\rho -\rho
\hat{a}_i\hat{a}_i^\dagger)\bigg\},
\end{aligned}
\end{equation}
where $\gamma_{i}$'s are the decay constants, and $N_{i}$'s 
represent the mean photon number of the individual baths.
Equation~\eqref{master_eqn_local} can be used to find the time
evolution of variances of quadrature operators, and hence,
the evolution of the covariance matrix.
%%%%%%%%%%%%%%%%%%%%%%%%%%%%%%%%%%%%%%%%%%%%%%%%%%%%%%%
\begin{figure}
\centering
\includegraphics[scale=1]{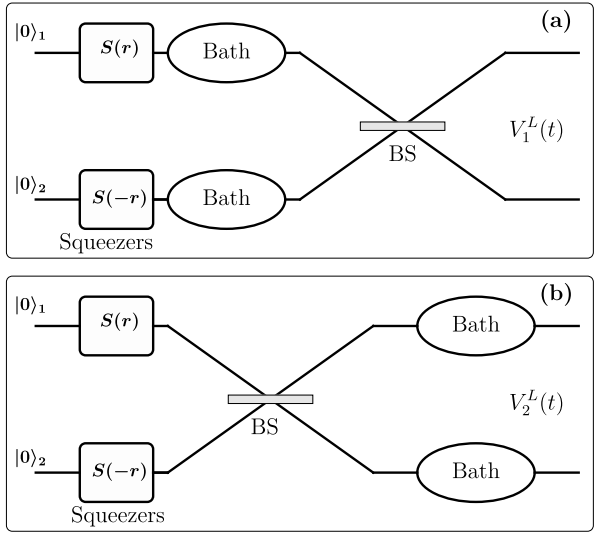} 
\caption{\label{fig:local_case}  Schematic representation of
the dissipation under local thermal baths. (a) Two-mode 
separable squeezed state is let to evolve under local
thermal baths and after a time $t$, the two modes are mixed
using a  $50:50$ beam splitter. 
(b)   Two-mode separable squeezed state is first mixed using
beam splitter and then it  is let to evolve under local
thermal baths. }
\end{figure}
%%%%%%%%%%%%%%%%%%%%%
\begin{figure}
\centering
\includegraphics[scale=1]{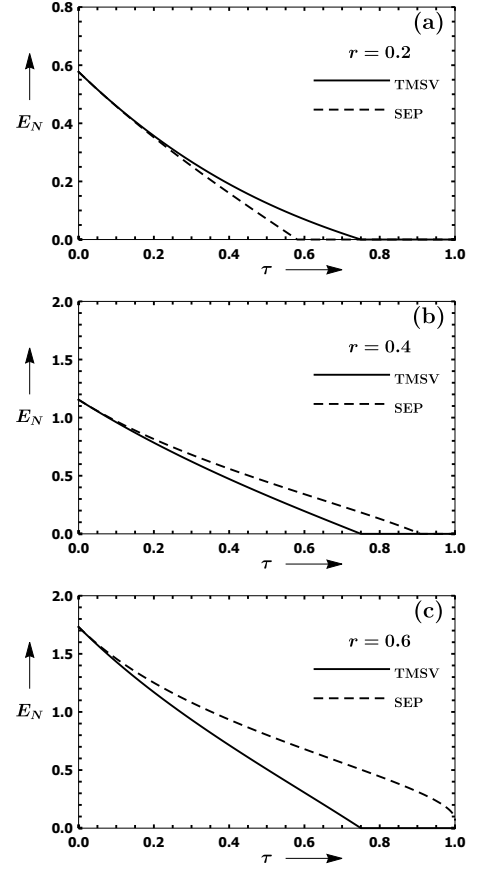} 
\caption{\label{fig:singlebath} Logarithmic negativity $E_N$
of two-mode Gaussian state as a function of dimensionless
time $\tau (=1-e^{-2\gamma t})$.  Only one of the two modes
is let to evolve under the presence of a local bath.  The
mean number of photons in the bath is taken to be $N=4$. (a)
For values of the initial squeezing parameter such that
$|r|<r_t^L<r_c^L$~[(\ref{rcritical}) and
(\ref{rtransition})], entanglement is more robust than
squeezing. (b) For $r_t^L<|r|<r_c^L$, squeezing is more
robust than entanglement. (c) For $|r|>r_c^L>r_t^L$,
squeezing is more robust than entanglement and the two-mode
separable case always remains entangled.}
\end{figure}
%%%%%%%%%%%%%%%%%%%%%%%%%%%%%%%%%%%%%%%%%%%%%%%%%%%%%%

\par
\noindent
\textbf{Case 1:} We  consider the case where each mode of
the two-mode separable squeezed state interacts with
distinct local thermal baths.  The schematic diagram is
shown in Fig.~\ref{fig:local_case}(a).
The covariance matrix of the initial state at time $t=0$ is
given by $V_1 (0)$~\eqref{initial_V_case1}.  Using the
master equation~\eqref{master_eqn_local}, we obtain the
covariance matrix after an interaction with the bath for time
$t$ as~\cite{olivares-2012}
\begin{equation}\label{int1_V_local_case1}
V_1(t)=X(t)V_{1}(0)X(t)^{T} + \frac{1}{2} Y(t),
\end{equation}
where $X(t)$ and $Y(t)$ are $4\times4$ diagonal matrices
given by
\begin{equation}
\begin{aligned}\label{X_and_Y}
X(t) &= \begin{pmatrix}(1-\tau_{1})^{1/4}\, \mathbb{1}_{2}& 0 \\
0 & (1-\tau_{2})^{1/4}\,\mathbb{1}_{2}
\end{pmatrix},\\
Y(t) &= \begin{pmatrix}
(\frac{N_{1}}{2}+1)\tau_{1}\, \mathbb{1}_{2}& 0 \\
0 & (\frac{N_{2}}{2}+1)\tau_{2}\,\mathbb{1}_{2}
\end{pmatrix},
\end{aligned}
\end{equation}
where $\tau_{1} = 1 - e^{-2\gamma_{1} t}$ and $\tau_{2} = 1
- e^{-2\gamma_{2} t}$ 
are dimensionless time parameters. We note that while $t$
goes from $0$ to $\infty$, $\tau_{i}$ goes from $0$ to $1$. 
The final covariance matrix after passing both the modes
through a 50:50 beam splitter~\eqref{beamsplitter} is given by
\begin{equation}\label{final_V_local_case1}
V^{L}_{1}(t) = B_{12}\left(\frac{\pi}{4}\right)\Big{[}
X(t)V_{1}(0)X(t)^{T} + 
 \frac{1}{2}Y(t)\Big{]}B_{12}\left(\frac{-\pi}{4}\right),
\end{equation}
where the superscript L over $V_{1}(t)$ stands for the local
bath, and $t$ represents the time duration of the
system-bath interaction. 
%%%%%%%%%%%%%%%%%%%%%%%%%%%%%%%%%%%%%%%%%%%%%%%
\par
\noindent
\textbf{Case 2: } We consider the case where each mode of
the TMSV state interacts with a distinct local thermal bath. 
The schematic is depicted in Fig.~\ref{fig:local_case}(b). 
The initial covariance matrix of the TMSV state is given by
Eq.~\eqref{initial_V_case2}.
Using the master equation~\eqref{master_eqn_local}, the
covariance matrix at time $t$ is evaluated as
\begin{equation}\label{final_V_local_case2}
V^{L}_{2}(t) = X(t)V_{2}(0)X(t)^{T} +  \frac{1}{2} Y(t).
\end{equation}

From Eqs.~(\ref{final_V_local_case1}) and
(\ref{final_V_local_case2}),  various conclusion can be drawn. 
We consider the following two special cases of symmetric and 
asymmetric interaction of  the local baths with the system:
\par
\noindent
\textbf{Symmetric interaction:} Consider the case where
$g_{1} = g_{2} = g$ and both baths are at same temperature
so that $\gamma_{1} = \gamma_{2} = \gamma$ and $N_{1} =
N_{2} = N$.  Thus the two local baths are identical. This
leads to the same final covariance matrix for two-mode
separable squeezed state~\eqref{final_V_local_case1} and
TMSV state~\eqref{final_V_local_case2}. Therefore both
the resources, squeezing and entanglement, are equally
sensitive to decoherence when two identical local thermal
baths act on each mode of the system.
\par
\noindent
\textbf{Extreme asymmetric interaction:} Consider $g_{1} =
g$ and $g_{2} = 0$, and therefore $\gamma_{1} = \gamma, \,
\gamma_{2} = 0$, and $N_{1} = N,\,N_{2}=0$.  Thus, only the first
mode of the system  interacts with the thermal bath.  The
covariance matrix for two-mode separable squeezed state~\eqref{final_V_local_case1}
 and TMSV state~\eqref{final_V_local_case2} are not the same in this
situation, and hence we expect different rates of decay for
logarithmic negativity of the final state. 

Using Eq.~\eqref{simon_criterion}, the condition on the 
initial squeezing parameter $r$, such that, the two-mode
separable squeezed state never becomes disentangled is given
by
\begin{equation}\label{rcritical}
|r| > r_c^L = \frac{1}{2}\left[\text{ln}\left( 1+\frac{N}{2}\right)\right].
\end{equation}
However, for values of $|r|\leq r_{c}^L$, entanglement dies
out for interaction times longer than
\begin{equation}\label{critical_time_sep_local}
\tau_a =
\frac{8\,e^{4|r|}[2+N-2\,\cosh(2|r|)]\,\sinh(2|r|)}{[(2+N)e^{2|r|}-2]^2}.
\end{equation}
On the other hand, the time of disentanglement for TMSV
state evaluates to 
\begin{equation}\label{critical_time_tmsv_local}
\tau_b = \frac{8(2+N)}{(4+N)^{2}},
\end{equation}
which is independent of the initial squeezing parameter $r$.
Further, $\lim\limits_{N\to 0} \tau_b$=1, which means the
entanglement of TMSV state survives indefinitely for a zero
temperature bath. In general,
Eqs.~\eqref{critical_time_sep_local} and
\eqref{critical_time_tmsv_local} imply the existence of finite
disentanglement time under the aforementioned conditions for
two-mode separable as well as TMSV states.  This
corresponds to the phenomenon of entanglement sudden death.

Further, if the initial squeezing parameter $r$ is such
that, $|r|$ is less than a certain value $r_t^L$, then it is
a better strategy to store the resource in the form
entanglement, otherwise it is better to store resources in
form of squeezing.  The expression for $r_t^L$ can be
evaluated by equating $\tau_a$ and $\tau_b$ and is given by
\begin{equation}\label{rtransition}
r_{t}^L = \frac{1}{2}\Big{[}\text{ln}
\Big{(}\frac{2+N+\sqrt{2(2+4N+N^2)}}{4+N}\Big{)}\Big{]}.
\end{equation}
We plot the  logarithmic negativity for different values of
initial squeezing parameter in Fig.~\ref{fig:singlebath}.
For $N=4$, the numerical values of $r_c^L$ and $r_t^L$ turn
out to be $r_c^L =0.55 $ and $r_t^L = 0.29 $.
Therefore,  for $r=0.20 < r_t^L<r_c^L$, we observe
that entanglement is better resource than squeezing.
For $r_t^L< r=0.40 < r_c^L $, 
we see that squeezing is better resource than entanglement.
Finally, for $r=0.60> r_c^L>r_t^L$, squeezing is better resource
than entanglement and the two-mode separable case always stays entangled.
%%%%%%%%%%%%%%%%%%%%%%%%%%%%%%%%%%%%%%%%%%%%
\subsection{Evolution under global Bath}
\label{resultsglobal}
In this section, we consider the scenario where the two
modes are coupled to a common thermal reservoir, which we
refer to as the global bath. The interaction Hamiltonian
$\hat{H}_{SB}$ is given by
\begin{equation}\label{int_hamiltonian_global}
\hat{H}_{SB} =
g\sum_{i=1,2}\Big{(}\hat{a}_{i}\sum^{\infty}_{k=1}\hat{b}^{\dagger}_{k}
 + \hat{a}^{\dagger}_{i}\sum^{\infty}_{k=1}\hat{b}_{k}\Big{)},
\end{equation}
where $\hat{b}_k$ and $\hat{b}^{\dagger}_k$ are annihilation
and creation operators of the $k^{th}$ mode of the
reservoir, and $g$ is the coupling constant between the
system and the environment. Under Markovian assumption, we
can write the master equation for the evolution of system
density operator $\rho$ as
\begin{equation}\label{master_eqn_global}
\begin{aligned}
\frac{\partial}{\partial t} \rho = &\frac{\gamma}{2} \bigg\{
\sum_{i=1,2} (N+1)(2 \hat{a}_i \rho
\hat{a}_i^{\dagger}-\hat{a}_i^{\dagger}\hat{a}_i\rho -\rho
\hat{a}_i^\dagger\hat{a}_i)\\
& +N(2 \hat{a}_i^\dagger \rho
\hat{a}_i-\hat{a}_i\hat{a}_i^{\dagger}\rho -\rho
\hat{a}_i\hat{a}_i^\dagger)\\
& + \sum_{j\neq i=1,2} (N+1)(2 \hat{a}_i \rho
\hat{a}_j^{\dagger}-\hat{a}_j^{\dagger}\hat{a}_i\rho -\rho
\hat{a}_j^\dagger\hat{a}_i)\\
& +N(2 \hat{a}_j^\dagger \rho
\hat{a}_i-\hat{a}_i\hat{a}_j^{\dagger}\rho -\rho
\hat{a}_i\hat{a}_j^\dagger)
\bigg\},
\end{aligned}
\end{equation}
where $\gamma$ is the decay constant, and $N$ represents the
mean photon number of the bath.
Equation~\eqref{master_eqn_global} can be used to find the
time evolution of the variances of quadrature operators, and
hence the time evolution of the covariance matrix.
%%%%%%%%%%%%%%%%%%%%%%%%%%%%%%%%%%%%%%%%%%%%%%%%%%%%%%
\par
\begin{figure}
\centering
\includegraphics[scale=1]{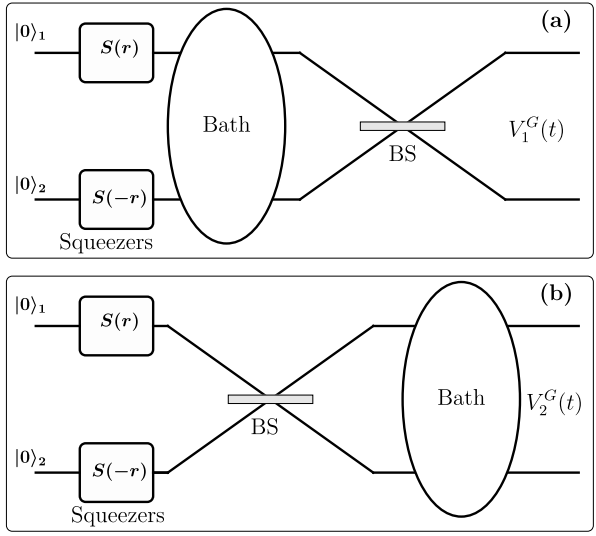} 
\caption{\label{fig:global_case} Schematic representation 
of the dissipation under a global bath. (a) The two-mode 
separable squeezed state is let to evolve under the global thermal 
bath, and after a time $t$, the two modes are mixed using a
$50:50$ beam splitter. 
(b) The two-mode separable squeezed state is first mixed 
using a  $50:50$ beam splitter, and then it  is let to
evolve under the global thermal bath.}
\end{figure}

\noindent
\textbf{Case 1: } We  consider the case where the two-mode
separable squeezed state interacts with a global bath. The
schematic diagram is shown in Fig.~\ref{fig:global_case}(a). 
The initial covariance matrix of the two-mode separable
squeezed  state is given by Eq.~\eqref{initial_V_case1}.
Using Eq.~\eqref{master_eqn_global}, the covariance matrix
after an interaction for time $t$ with the bath is evaluated
to be
\begin{equation}\label{intermediate_V_global_case1}
\frac{1}{4}\left(\begin{array}{cccc}
\sigma_{1}(t)& 0 & \sigma_{3}(t) & 0\\
0 & \sigma_{2}(t) & 0 & \sigma_{3}(t)\\
\sigma_{3}(t) & 0 & \sigma_{2}(t) & 0\\
0 & \sigma_{3}(t) & 0 & \sigma_{1}(t)\end{array} \right),
\end{equation}
where
\begin{equation}\label{intermediate_V_global_case1_notations}
\begin{split}
\sigma_{1}(t) & = (2N+1)\tau - \cosh(2r)(\tau - 2) -
2\,\sinh(2r)\sqrt{1-\tau},\\
\sigma_{2}(t) & = (2N+1)\tau - \cosh(2r)(\tau - 2) +
2\,\sinh(2r)\sqrt{1-\tau},\\
\sigma_{3}(t) & = (2N+1-\cosh(2r))\tau,\\
\end{split}
\end{equation}
where $\tau = 1 - e^{-2\gamma t}$. 
We now pass both the modes through a 50:50 beam splitter,
and the resultant 
covariance matrix is given by
\begin{equation}\label{final_V_global_case1}
V^{G}_{\text{1}}(t)=\frac{1}{2}\left(\begin{array}{cccc}
\sigma'_{1}(t)& 0 & \sigma'_{3}(t) & 0\\
0 & \sigma'_{1}(t) & 0 & -\sigma'_{3}(t)\\
\sigma'_{3}(t) & 0 & \sigma'_{2}(t) & 0\\
0 & -\sigma'_{3}(t) & 0 & \sigma'_{2}(t)\end{array} \right),
\end{equation}
where
\begin{equation}\label{final_V_global_case1_notations}
\begin{split}
\sigma'_{1}(t) & = (2N+1)\tau - \cosh(2r)(\tau - 1),\\
\sigma'_{2}(t) & = \cosh(2r),\\
\sigma'_{3}(t) & =  \sinh(2r)\sqrt{1-\tau}.
\end{split}
\end{equation}
%%%%%%%%%%%%%%%%%%%%%%%%%%%%%%%%%%%%%%%%%%%%
%%%%%%
\begin{figure}[htbp] 
\centering
\includegraphics[scale=1]{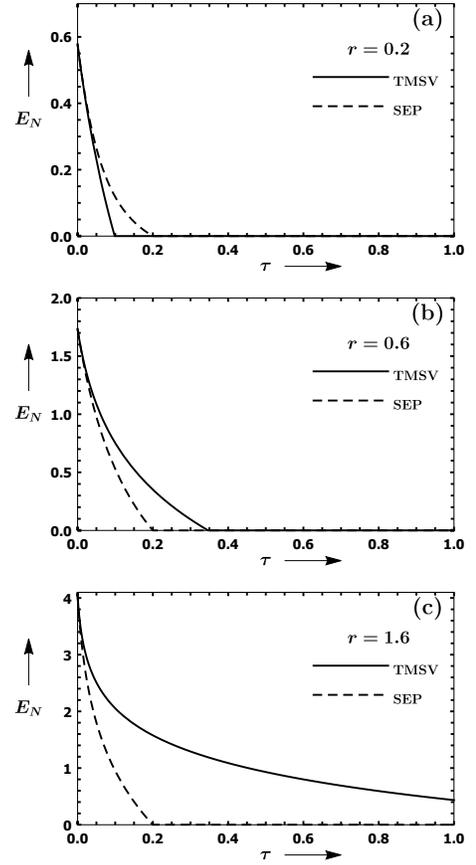} 
\caption{\label{fig:lagarithmic_negitivity_plots}
Logarithmic negativity
$E_N$  as a function of dimensionless time $\tau
(=1-e^{-2\gamma t})$.  The system is let to evolve under the
presence of a global thermal bath.  The mean number of photons
in the bath has been taken to be $N=4$.  
(a) For values of the initial
squeezing parameter such that $|r|<r_t^G<r_c^G$~[(\ref{rcriticalg}) and 
(\ref{rtransitionglobal})], squeezing is more robust than
entanglement.  (b) For $r_t^G<|r|<r_c^G$, entanglement is more
robust than squeezing. (c) For $|r|>r_c^G>r_t^G$, entanglement is more
robust than squeezing and the TMSV state always remains entangled.}
\end{figure}

%%%%%%%%%%%%%%%%%%%%%%%%%%%%%%%%%%%%%%%%%%%%%%%%%%%5
\par
\noindent
\textbf{Case 2: } We consider the case where the TMSV state
interacts with a global bath.   The
schematic is presented in Fig.~\ref{fig:global_case}(b). 
The initial covariance
matrix of the TMSV state is given by
Eq.~\eqref{initial_V_case2}.  The  covariance matrix after
an interaction for time $t$ with the bath is evaluated to be
\begin{equation}\label{final_V_global_case2}
V^{G}_{2}(t)=\frac{1}{2}\left(\begin{array}{cccc}
\delta_{1}(t)& 0 & \delta_{3}(t) & 0\\
0 & \delta_{2}(t) & 0 & \delta_{4}(t)\\
\delta_{3}(t) & 0 & \delta_{1}(t) & 0\\
0 & \delta_{4}(t) & 0 & \delta_{2}(t)\end{array} \right),
\end{equation}
where
\begin{equation}\label{final_V_global_case2_notations}
\begin{split}
\delta_{1}(t) & = \frac{1}{2}(2N+1-e^{2r})\tau + \cosh (2r),\\
\delta_{2}(t) & = \frac{1}{2}(2N+1-e^{-2r})\tau + \cosh (2r),\\
\delta_{3}(t) & = \frac{1}{2}(2N+1-e^{2r})\tau + \sinh (2r),\\
\delta_{4}(t) & = \frac{1}{2}(2N+1-e^{-2r})\tau - \sinh (2r).
\end{split}
\end{equation}
%%%%%%%%%%%%%%%%%%%%%%%%%%%%%%%%%%%%%%%%%%%
The evolved covariance matrix for the two-mode separable
squeezed state~\eqref{final_V_global_case1} and the TMSV
state~\eqref{final_V_global_case2} are not the same in
this situation, and hence we expect different rates of decay
for logarithmic negativity.  For the TMSV state, there exists a
critical value of initial squeezing parameter $r_c^G$,  above
which the  entanglement never becomes zero~\cite{prauzner-2004}:
\begin{equation}\label{rcriticalg} |r| >r_c^G=
\frac{1}{2}[\text{ln}(2N+1)].
\end{equation}
For values of $|r|$ less than $r_c^G$, entanglement sudden
death occurs after a time
\begin{equation}
\tau_c = \frac{2\,\sinh(2|r|)}{1-e^{-2|r|}+2N}.
\end{equation}
It is also observed that for two-mode separable squeezed
state, entanglement
 always becomes zero at a particular value of time
irrespective of the initial  squeezing of the state:
\begin{equation}
\tau_d = \frac{1}{1+N}.
\end{equation}
Thus entanglement sudden death always occurs for two-mode
separable squeezed state except for zero temperature bath.

If the initial squeezing parameter $r$ is such that, $|r|$
is less than a certain value $r_t^G$, then it is a better
strategy to store the quantum resource in the form squeezing,
otherwise it is better to store the quantumness in the form of
entanglement. The expression for $r_t^G$ can be evaluated
by equating $\tau_c$ and $\tau_d$ and is given by
 \begin{equation}\label{rtransitionglobal}
 r_{t}^G = \frac{1}{2}\Big{[}\text{ln}
\Big{(}\frac{1+2N+\sqrt{1+8N+8N^2}}{2(1+N)}\Big{)}\Big{]}.
\end{equation}
We have shown the plots of   logarithmic negativity for
different values of initial squeezing parameter in Fig.~\ref{fig:lagarithmic_negitivity_plots}.
For $N=4$, the numerical values of $r_c^G$ and $r_t^G$ turn out to be $r_c^G =1.10 $ and $r_t^G = 0.39 $.
Therefore, in corroboration with analytical results,
 we observe that for $r=0.20 < r_t^G<r_c^G$, 
squeezing is better resource than entanglement.
Further,  for $r_t^G< r=0.60 < r_c^G$, entanglement 
is better resource than squeezing.
Finally, for $r=1.60> r_c^G>r_t^G$, entanglement 
is better resource than squeezing; however, the TMSV state always remains entangled.
We have summarized the results of this section in
Table~\ref{table1}.

\begin{table}[H]
\caption{\label{table1}
Better resource for storage of quantumness: squeezing or
entanglement}
\renewcommand{\arraystretch}{1.5}
\begin{tabular}{|p{3.1cm}|c|c| } 
\hline \hline
Environment Nature & $r$ Range &Result\\
\hline \hline
Identical local baths  & & Squeezing $\equiv$ Entanglement \\ 
\hline
Single local bath &  $|r| < r_t^L$ & Squeezing $<$ Entanglement \\
&  $|r| > r_t^L$& Squeezing $>$ Entanglement \\ \hline
Global bath  & $|r| < r_t^G$ & Squeezing $>$ Entanglement \\
&  $|r| > r_t^G$& Squeezing $<$ Entanglement \\ 
\hline \hline
\end{tabular}
\end{table}
%%%%%%%%%%
\section{Conclusion}
\label{conc}
In this paper, we compared the robustness of squeezing and
entanglement resources in two-mode Gaussian states evolving in
different noisy dissipative environments. To this end, we
considered two different cases where in the first case, two
mode separable squeezed states were allowed to evolve in the
dissipative environment, and then were entangled  using
passive operations. In the second case, the squeezed modes
were first entangled using passive operations and then were
evolved in the dissipative environment.  The results show
that the robustness of squeezing and entanglement depends on
the initial squeezing of the state and the nature of the
dissipative environment. We also observe entanglement sudden
death in specific cases.  The fact that interconversion of
squeezing and entanglement can be made by using passive
optical elements like beam splitters, wave plates, phase
shifters and mirrors,  makes it convenient to save the
resource in one or other form depending on the situation.
One of the directions that we are pursuing is to generalise
this work by considering more general environment models and
 non-Gaussian states which will have implications for
recent CV based key distribution~\cite{Zubairy-pra-2020} and
quantum
teleportation~\cite{Hyunchul-pra-2013,Zubairy-pra-2017}
protocols.

%%%%%%%%%%%%%%%%%%%%%%%%%%%%%%%%%
\section*{Acknowledgement}
   A and C.K. acknowledge the financial support from {\sf
	DST/ICPS/QuST/Theme-1/2019/General} Project number {\sf
	Q-68}. 
%%\bibliography{ref}
\end{document}